\documentclass[aps,pra,nofootinbib,preprint,amsmath,amssymb,floatfix]{revtex4}
\begin{document}

\section*{Reply to ''Comment on ''Analysis of recent interpretations of the Abraham-Minkowski problem''''}

{\bf Abstract}

The  Comment of M. Partanen and J. Tulkki [Phys. Rev. A {\bf 100}, 017801 (2019)] claims that my criticism expressed in Phys. Rev. A {\bf 98}, 043847 (2018) of the earlier paper of Partanen {\it et al.} [Phys. Rev. A {\bf 95}, 063850 (2017)] was   incorrect. Now, there  are essentially three points involved here: (1) the first one regards the physical interpretation of the radiation pressure experiment of A. Kundu {\it et al.} [Sci. Rep. {\bf 7}, 42538 (2017)]. My mathematical analysis of this situation was certainly simplistic, but yet it was able to illustrate the main property of the experiment, namely that it showed the  action from the radiation forces  on the dielectric boundaries, and  from the Lorentz force in the interior, but it had not any relationship to the Abraham-Minkowski momentum problem as was originally stated by the investigators. (2) The second and most important point was my emphasis on the fact that in an electromagnetic pulse in a medium there cannot be an accompanying mechanical energy density of the same order of magnitude as the electromagnetic energy density itself. The electromagnetic wave carries with it a mechanical {\it momentum}, but not a mechanical {\it energy}
 (the latter being of second order in the particle velocity). In the present note I illustrate this point by a simple numerical analysis. (3) When going to a relativistic formulation of the electromagnetic theory in matter, care must be taken to secure that  fundamental conditions from field theory are satisfied. In particular, one cannot in general take the electromagnetic total energy and momentum of a radiation pulse to constitute a four-vector; such a property holds only if the electromagnetic energy-momentum tensor is divergence-free. For the Minkowski tensor this condition is satisfied, whereas for the Abraham tensor it is not. As a conclusion, my earlier standpoint on the Abraham-Minkowski problem remains unchanged.

 ------------------------------------------------------------------------------------------------------------

\bigskip
As mentioned in the Abstract above, the  Comment by Partanen and Tulkki (PT) \cite{partanen19} on my recent paper \cite{brevik18} consists essentially of the following points:

\noindent 1. First, my analysis of the radiation pressure experiment of Kundu et al. \cite{kundu17} was considered. Kundu et al. claimed this experiment to show the correctness of the Abraham photon momentum in matter, in contrast to the Minkowski momentum. One of my purposes in the analysis in \cite{brevik18} was to point out that this claim was not right. What the experiment does, is to show the combined effect from the surface forces (common for the Abraham and Minkowski alternatives) and the Lorentz forces on charges in the interior. The Abraham momentum does not come into play here at all. PT claim that my elastic model was simplistic, not able to describe the irreversible character of the experiment. That statement is in itself right. My intention was however not to describe the real experiment in detail, but to adopt a simple elastic model that was sufficient to explain the basic physics. The physical effect is  solely due to the combined action of electromagnetic forces in the boundary regions, and in the interior.

\noindent 2. The second, and a more important point,  was the analysis of an optical pulse traveling in a resting medium. The core of the problem is shown in  the Comment's expression (10) for the energy-momentum tensor. In this context it is convenient to consider an example. Let a cw laser beam $E_{inc}=E_0e^{i(kz-\omega t)}$ be propagating in the $z$ direction in water. Assume power $P=1~$watt. With a cross section of $1~$mm$^2$ this corresponds to a Poynting vector ${\bf S}={\bf E\times H}$ of magnitude
 $S=1~$MW/m$^2$. The electromagnetic field energy density is
 \begin{equation}
 W=\frac{1}{2}{\bf (E\cdot D+H\cdot B)},
 \label{1}
 \end{equation}
which amounts to $W=Sn/c=  4.44~ \rm{mJ/m}^3$ when $n=1.33$ is the refractive index.
 The mechanical momentum density given to the medium from the leading edge of the pulse, here called ${\bf g}^{\rm mech}$, is
\begin{equation}
{\bf g}^{\rm{ mech}}
= \frac{n^2-1}{c^2}({\bf E\times H})
=\frac{n^2-1}{c^2}{\bf S}, \label{2}
\end{equation}
in accordance with Eq.~(8) in PT, and is correct. (The existence of this momentum density  was demonstrated experimentally by R. V. Jones and others a long time ago \cite{jones54,jones78}.) Now put the expression (\ref{2}) equal to $\Delta p\cdot N$, with $\Delta p$  the impulse given to each molecule and $N$ the particle density. Taking the molecule  mass $m$ to be $18u$ with $u=1.66\times 10^{-27}~$kg the atomic mass unit, we get $N=3.35\times 10^{28}~$m$^{-3}$, and $\Delta p=2.55\times 10^{-40}~$N\,s. With $\Delta p=mv$ we thus obtain for the particle velocity $v=0.85 \times 10^{-14}~\rm{m/s}$.
It corresponds to a kinetic mechanical energy density
\begin{equation}
E_{\rm kin}=\frac{1}{2}\rho v^2=3.6\times 10^{-25}~\rm{J/m^3}, \label{3}
\end{equation}
with $\rho$ the density of water.

Now consider Eq.~(7) in PT, where the energy density for a propagating wave, called $W_{\rm MDW}^{(L)}$, is given as
\begin{equation}
W_{\rm MDW}^{(L)}= (n^2-1)W, \label{4}
\end{equation}
where  $W$ is the same as in  Eq.~(\ref{1}) above. Expression (\ref{4}) means that there should be an extra energy term of order $W$, in addition to the electromagnetic energy density $W$, in the propagating wave.  Where can this extra term come from? Obviously not from the immensely smaller kinetic energy density $E_{\rm kin}$ in Eq.~(\ref{3}). I cannot see that the PT energy density expression, Eq.~(\ref{4}) above, can be  correct.

There is in addition another special property of the PT formalism that causes concern: according to their energy-momentum tensor (10)  the energy flux density  should be equal to $n^2\bf({E\times H})$, instead of the usual Poynting vector $\bf({E\times H})$. Such a change of the expression for the energy flux density, if correct, would simply revolutionize classical electromagnetic theory. It cannot be right.

\noindent 3.  In their analysis, PT made   use of relativistic arguments, involving the rest mass energy $\delta m c^2$. This may be an unfortunate approach, as it tends   to obscure  the argument. Physically, the present problem involves only weak mechanical forces in a resting medium, and has very little to do with relativity, although it is true that  relativity formally turns up in connection with the transformations of energy and momentum that PT made use of. From a general perspective it is  known that in order for energy and momentum to transform as the components of a four-vector, the corresponding four-divergence of the energy-momentum tensor has to be zero. This has been shown by M{\o}ller and others \cite{moller72}. The Minkowski tensor satisfies this criterion, as its energy-momentum tensor (superscript M) obeys the equation $\partial_\nu T_{\mu\nu}^M=0$ in view of Maxwell's equations.

\noindent 4. Remark on Appendix A in the Comment:  Equation (A1) is the same as the Minkowski/Abraham conservation equation  for energy, when multiplied with the constant $n^2$. Equation (A2) is precisely  the Minkowski conservation equation for momentum. Evidently, the four-divergence of the energy-momentum tensor $T_{MP}$ in (10)
is therefore equal to zero.

In conclusion, I have to uphold my earlier standpoint on the Abraham-Minkowski problem as expressed in  Ref.~\cite{brevik18}. Especially, this concerns  Point 2 above.

\newpage
 Iver Brevik

Department of Energy and Process Engineering

Norwegian University of Science and Technology,

7491 Trondheim, Norway

\end{document}